# Ac-magnetic susceptibility in the peak-effect region of Nb$_3$Sn

O. Bossen[1], A. Schilling[1], and N. Toyota[2]

[1]Physik-Institut University of Zurich, Winterthurerstrasse 190, CH-8057 Zurich, Switzerland

[2]Physics Department, Graduate School of Science, Tohoku University, 980-8571 Sendai, Japan

We performed a systematic study of the ac magnetic-susceptibility on a Nb$_3$Sn single crystal which displays a strong peak effect near the upper critical field $H_{c2}$. In external magnetic fields above $\mu_0 H \approx 3\,\text{T}$, the peak effect manifests itself in a single, distinct peak in the real part $\chi'(T)$ of the ac susceptibility as a function of temperature $T$, the size of which continuously increases with increasing magnetic field $H$. In the imaginary part $\chi''(T)$ of the ac susceptibility, on the other hand, a single peak initially grows with increasing $H$ up to a well-defined value, and then splits into two sharp peaks which separate when $H$ is further increased. We explain this surprising behavior by a flux-creep model and taking into account the enhancement of the critical-current density in the peak-effect region near $T_c$ in which Bean's critical-state model seems to apply. Outside this region, the crystal is clearly in a flux-creep regime with finite creep exponent $n$.

## I. Introduction

The peak effect in superconductors was first discovered in 1961, when both Berlincourt *et al.* and LeBlanc and Little found a strong increase of the critical-current density near the upper critical field $H_{c2}$ [1,2]. Pippard explained this behaviour by a "softening" of the vortex lattice at elevated temperatures [3]. With increasing temperature $T$, the shear modulus vanishes as $(T_c-T)^2$ near the critical temperature $T_c(H)$, while the pinning interactions vary linearly with $(T-T_c)$, thereby leading to an enhanced vortex pinning in a certain narrow range of temperatures near $T_c$. In 2006 Adesso *et al.* [4] measured for the first time a corresponding peak effect in the superconductor $Nb_3Sn$. This work was performed on the same single crystal that is under study here, and it has attracted much interest because of the importance to achieve high critical-current densities in $Nb_3Sn$ for technical applications. Further studies on the nature of the peak effect in this crystal were communicated by Lortz *et al.* [5] who studied the dc magnetization and the specific heat in the peak effect region, and by Reibelt *et al.* [6] who showed that an additional small ac magnetic "shaking" field can reveal the peak effect in resistivity measurements as well.

## II. Modelling the ac susceptibility

In ac magnetic-susceptibility measurements a sample is subject to a small oscillating magnetic excitation field with amplitude $H_{ac}$ and frequency $f$, and the magnetic response of the sample is determined from the induced e.m.f. in a secondary pick-up coil. Both the in-phase response (real part) and the out-of-phase response (imaginary part) of the recorded signal are of interest. Type I superconductors are in the Meissner state and therefore expel the applied magnetic field completely. The ac magnetic-susceptibility $\chi = \chi' + i\chi''$ then has no imaginary part (i.e. no dissipative losses), and the real part is $\chi' = -1$. Type II superconductors in the mixed state, on the other hand, do not expel the magnetic field completely, and they can show considerable magnetic hysteresis due to vortex pinning. Therefore, $|\chi'|<1$ and $\chi''>0$ as soon as $H>H_{c1}$, with $H_{c1}$ the lower critical field. The hysteretic losses are related to $\chi''$ via

$$P = \mu_0 \pi f H_{ac}^2 \chi'' V, \qquad (1)$$

where $P$ is the dissipated power and $V$ the sample volume [7].

### a) Resistive model

For a sample showing Ohmic behavior, the ac magnetic losses stem from the resistive losses of the induced electrical eddy currents. They reach a maximum when the skin penetration depth is of the order of the magnitude of the sample size. For an infinite slab of thickness $d$, resistivity $\rho$, and the probing ac magnetic field $H_{ac}(t)$ in parallel to the surface of the slab, the resulting ac magnetic-susceptibility is given by [8]

$$\chi' = \frac{\sinh u + \sin u}{u(\cosh u + \cos u)}, \qquad (2a)$$

$$\chi'' = \frac{\sinh u - \sin u}{u(\cosh u + \cos u)}, \qquad (2b)$$

$$\text{with} \quad u = \left(\frac{\mu_0 \omega d^2}{2\rho}\right)^{1/2}, \qquad (2c)$$

where $\omega = 2\pi f$ is the angular frequency of the ac magnetic field. Both $\chi'$ and $\chi''$ are uniquely determined by the dimensionless parameter $u$, are independent of the amplitude $H_{ac}$, and as such fulfill a universal relationship $\chi''(\chi')$ for all Ohmic slabs of this geometry.

**b) Critical-state model**

A successful model to quantitatively explain the magnetic hysteresis in type II superconductors was introduced by Bean in 1964 [9]. The model assumes that magnetic flux enters a superconductor from the outside, thereby inducing a shielding surface current with critical-current density $j_c$. When the external magnetic field is reduced or reversed, corresponding surface currents with magnitude $j_c$ but with opposite orientation develop from the surface. The corresponding magnetic susceptibility derived from this model, again for an infinite slab parallel to the ac excitation field, takes the frequency-independent form [10]

$$\chi' = -1 + \frac{H_{ac}}{j_c d}, \tag{3a}$$

$$\chi'' = \frac{4 H_{ac}}{3\pi j_c d}, \qquad H_{ac} \leq H^* \tag{3b}$$

and

$$\chi' = -\frac{j_c d}{4 H_{ac}}, \tag{3c}$$

$$\chi'' = \frac{j_c d}{\pi H_{ac}} - \frac{j_c^2 d^2}{3\pi H_{ac}^2}, \qquad H_{ac} > H^* \tag{3d}$$

with $H^* = j_c d / 2$. Using Eqs. (3a)-(3d) we can express $\chi''$ as a function of $\chi'$, thereby eliminating the explicit dependence on $j_c, d$ and $H_{ac}$,

$$\chi'' = \frac{4}{3\pi}(\chi' + 1), \qquad \chi' \leq -\frac{1}{2} \tag{4a}$$

$$\chi'' = -\frac{4}{\pi}\chi' - \frac{16}{3\pi}\chi'^2, \qquad \chi' > -\frac{1}{2} \tag{4b}$$

## c) Flux-creep model

Brandt [12] noticed that the resistive model and the Bean model discussed above can be interpolated using a flux-creep model with a non Ohmic power-law dependence of the electric field $E$ on the current density $j$, i.e.,

$$E \propto \left[ j / j_c(B) \right]^n, \qquad (5)$$

where $j_c(B)$ is the field-dependent critical-current density and $n$ the creep exponent. The Eq. (5) reproduces Ohms law for $n = 1$, and Bean's critical-state model corresponds to a sudden onset of resistive behaviour for $n \to \infty$. It turns out that for a given finite exponent $n$, $\chi''(\chi')$ again obeys a universal relationship, independent of excitation frequency $f$ and amplitude $H_{ac}$, since both quantities are contained in a dimensionless parameter that determines both $\chi'$ and $\chi''$ [11]. However, these susceptibilities can, in general, not be expressed in an analytical form for arbitrary $n$ and sample geometries, and we will therefore rely in the following on the calculations for bar-shaped samples from Ref. [11].

## III. Ac susceptibility in the peak-effect region of $Nb_3Sn$

Without any peak effect near $T_c$, the critical-current density $j_c(T)$ decreases monotonically with increasing temperature $T$ and vanishes at $T = T_c$. As a result, the real part $\chi'$ of the magnetic susceptibility also monotonically increases with $T$ before reaching zero at $T_c$ (see Fig. 1a, left panel). In the Bean model with a slab geometry, the imaginary part $\chi''$ has a maximum $\chi''_m = \frac{3}{4\pi} \approx 0.239$ for $\chi'_m = -0.375$ (Fig. 1a, right panel). For different geometries and finite creep exponents, different values $\chi''_m$ are assumed for slightly different values of $\chi''_m$ (e.g., $\chi''_m \approx 0.417$ for $\chi'_m \approx -0.417$ in the resistive model for a slab geometry, or

$\chi''_m \approx 0.32$ for $\chi'_m \approx -0.36$ as numerically estimated by Brandt for a finite rectangular bar with a height-to-width ratio of 1 and a creep exponent of $n = 3$ [12]).

If the sample displays a peak effect near $T_c$, however, the critical-current density increases sharply before dropping to zero at the transition to the normal state [1,2,3]. This causes a sudden decrease in $\chi'(T)$, i.e., a peak in $|\chi'(T)|$ (Figs. 1b and 1c). Depending on the magnitude of this peak which is determined by the pinning strength in the peak-effect region, the sample geometry and the probing ac magnetic field, this will either lead to a single peak in the imaginary part $\chi''(T)$ of the susceptibility if $|\chi'| \leq |\chi'_m|$ (Fig. 1b), or to a double-peak structure as soon as $|\chi'| > |\chi'_m|$ in the peak effect region (Fig. 1c). The height of these double peaks corresponds to $\chi''_m$ (Fig. 1c, left panel).

This is the main result of our paper. It remains qualitatively correct even if other models to calculate the ac susceptibility are used, as long as $\chi''$ is a single-valued function of $\chi'$ with a single maximum at some intermediate value of $\chi'_m$ as sketched in the right panel of Fig. 1a.

To test our scenario we have performed a systematic study of the ac magnetic susceptibility in external magnetic fields ranging from zero up to $\mu_0 H = 9$ T in steps of 1.5 T, with excitation amplitudes $\mu_0 H_{ac} = 0.5$ mT, 1.0 mT, 1.5 mT, and 17.0 mT, frequencies $f = 200$ Hz and $f = 1$ kHz, and in a temperature range between $T = 4$ K and $T_c = 18.2$ K. These measurements were done using the ACMS option of a Physical Properties Measurement System *(Quantum Design)*. The Nb$_3$Sn crystal was of rectangular shape with dimensions $\approx 3 \times 1.3 \times 0.4$ mm$^3$.

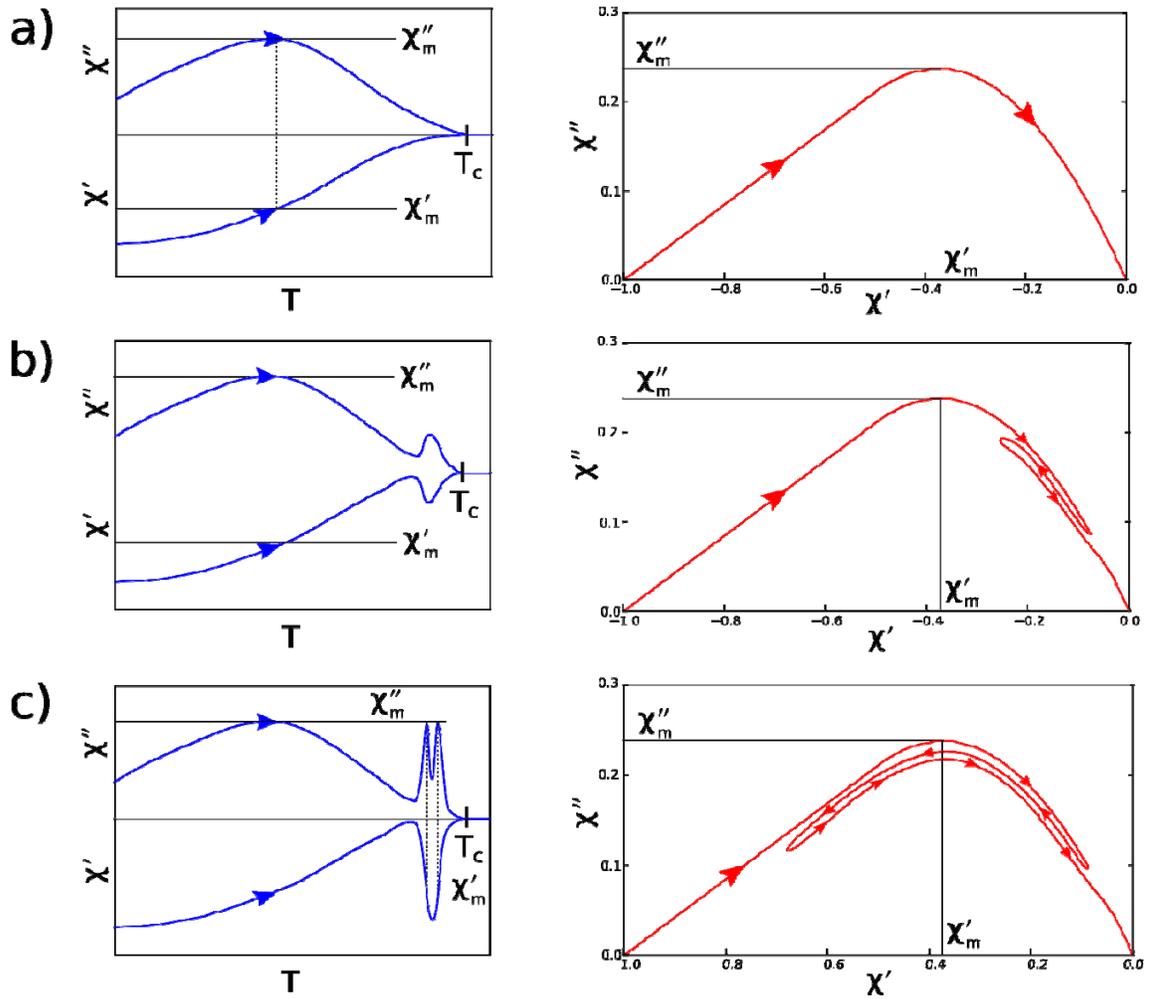

Fig. 1, left panels: Real and imaginary parts of the ac magnetic-susceptibility for a superconductor displaying a peak effect. Right panels: corresponding $\chi''$ vs. $\chi'$ representations. Arrows indicate an experiment with increasing temperature, for a): a type II superconductor without peak effect; b): a type II superconductor with a weak peak effect, producing single peaks in both real and imaginary parts of the ac susceptibility; c): a type II superconductor showing a strong peak effect, leading to a single peak in the real part but a double-peak structure in the imaginary part.

The measured complex ac-susceptibility $\chi_{meas}$-raw-data must, in principle, be corrected according to

$$\chi = \frac{\chi_{meas}}{1-D\chi_{meas}} \quad (6)$$

with $D$ the real-valued demagnetization factor, where both $\chi$ and $\chi_{meas}$ are complex numbers with real and imaginary parts $\chi'$ and $\chi''$, respectively [13]. The factor $D \approx 0.66$ for our crystal was obtained from corresponding ac-susceptibility measurements taken in zero external magnetic field at $T = 4$ K with $\mu_0 H_{ac} = 1.7$ mT and assuming complete magnetic-flux expulsion, i.e., $\chi' = -1$. This value for $D$ is in reasonable agreement with the geometry of the crystal for which we can estimate $D \approx 0.6 - 0.8$ as calculated for a rectangular prism [14]. Published detailed calculations of $\chi'$ and $\chi''$ values by Brandt [12] for various sample geometries and creep exponents $n$ had been normalized to $\chi' = -1$ in the limit $H_{ac} \to 0$, but rather using a constant factor instead of applying Eq. (6) [12,15]. To directly compare our measurements with these calculations, we had therefore to normalize our $\chi'$ and $\chi''$ data in the same way, i.e., by multiplying $\chi_{meas}$ with (1-$D$).

Representative susceptibility data of this study are shown in Figs. 2. Our results for the ac magnetic susceptibility show negligible dependence on frequency, and the real part agrees well with the results of Adesso *et al.* [4]. Above $\mu_0 H \approx 3\mathrm{T}$, a peak in $\chi'(T)$ starts to form close to $T_c$, indicating the increase of the critical-current density due to the peak effect (Figs. 2, left panels). A corresponding single peak in the imaginary part $\chi''(T)$ appears along with this peak in $\chi'(T)$ and grows with $H$ up to $\mu_0 H = 4.5\mathrm{T}$, beyond which it is indeed splitting into two peaks (left panels of Figs. 2b - 2d.) In the right panels of Fig. 2 we have plotted the corresponding $\chi''$ vs. $\chi'$ representations. To compare these data with theoretical predictions and to better visualize the sequence of the data points, we have plotted $\chi''$ vs. $\chi'$ for $\mu_0 H_{ac} = 1.5\,\text{mT}$ in a separate graph for clarity (Figs. 3a-d).

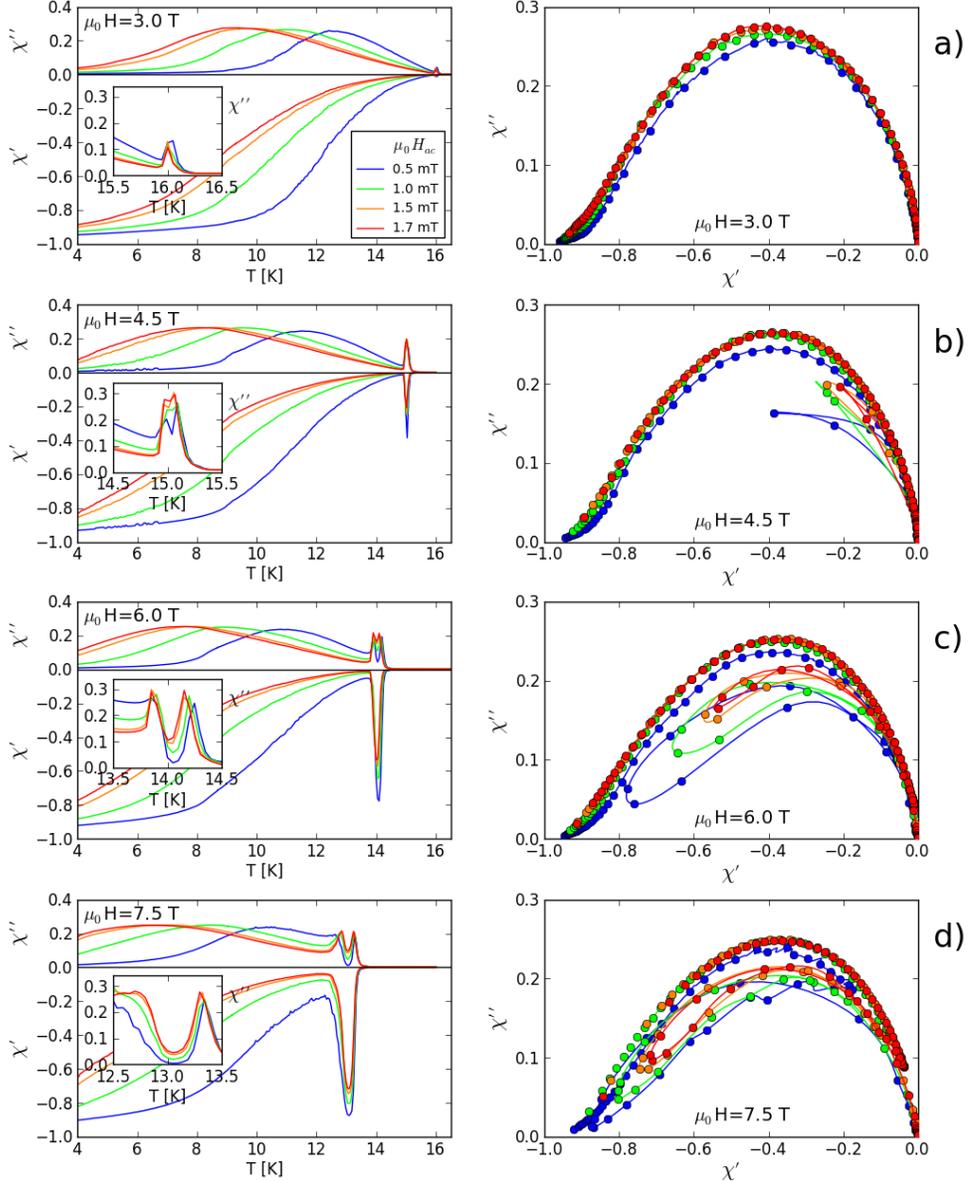

Fig. 2, left panels: Ac magnetic-susceptibility data of a Nb$_3$Sn single crystal. The magnetic field $\mu_0 H = 3.0\,\text{T}$ corresponds to case b) in Fig. 1, and the data taken in $\mu_0 H = 4.5\,\text{T}$ and above to case c) from Fig. 1. Right panels: Corresponding $\chi''(\chi')$ representations, together with interpolating spline fits to guide the eye.

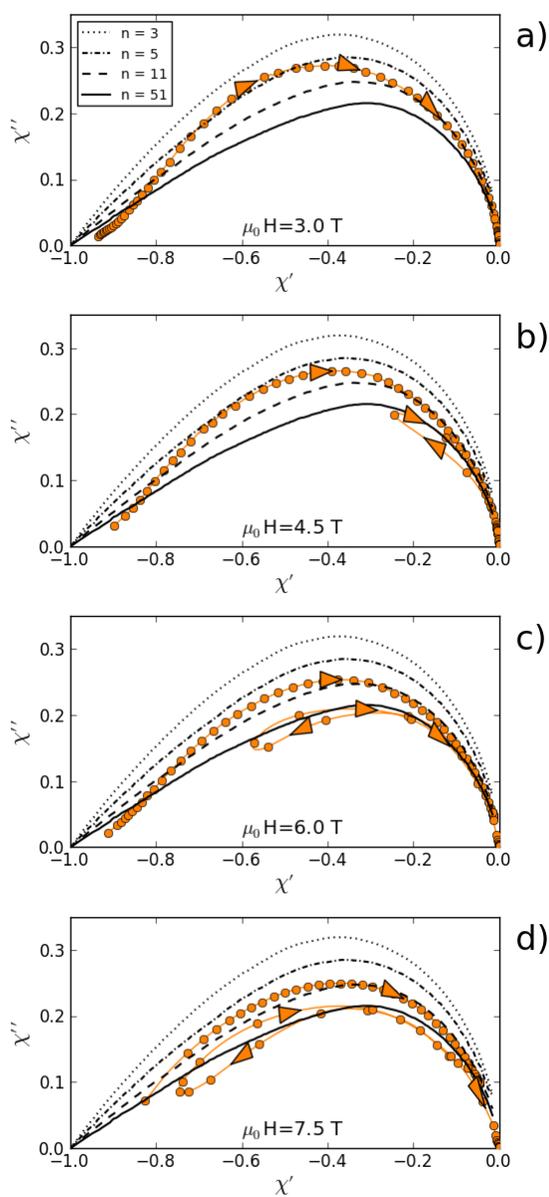

Fig 3: Theoretical expectations for $\chi''(\chi')$ from interpolations according to Brandt [12] for a finite-bar geometry with creep exponents ranging from $n = 3$ to $n = 51$. Only the data for $\mu_0 H_{ac} = 1.5\,\text{mT}$ are shown here for clarity, together with an interpolating spline fit with arrows to visualize the sequence of the data points.

The overall behavior of $\chi''(\chi')$ in the peak-effect region can be quite well explained by a universal $\chi''(\chi')$ relationship, particularly for $\mu_0 H = 3\,\text{T}$. In the narrow peak-effect region for $\mu_0 H = 4.5\,\text{T}$ and above, however, this relationship does not hold exactly, and $\chi''(\chi')$ becomes multi-valued (see right panels of Figs. 2, and Fig 3). We can attribute this behavior to the fact that the two regimes inside and outside the peak-effect region belong to two different categories, i.e. different creep exponents $n$. The $Nb_3Sn$ crystal must be in a flux-creep regime below the peak effect region, which is suggested by the investigations of Reibelt *et al.* on the same sample [6], in which the appearance of a finite measurable resistance for $\mu_0 H_{ac} > 0.3\,\text{mT}$ and $f > 1\,\text{kHz}$ was observed. Once the excitation was removed, the resistance dropped again to zero, a behavior which is characteristic for the flux-creep regime [16,17]. This interpretation is supported by the fact that $\chi''(\chi')$ outside the peak effect region reasonably well follows an interpolation by Brandt [12] for a finite-bar geometry with a creep exponent ranging from $n \approx 5$ in $\mu_0 H = 3$ T to $n \approx 11$ in $\mu_0 H = 7.5$ T (see Fig. 3, dash-dotted and dashed lines). In the peak-effect region, by contrast, the $\chi''(\chi')$ data are very close to the prediction of the Bean model with a large exponent $n \approx 51$ (see again Figs. 3, solid line [12]). We note here that similar multiple peaks in the pendulum data of several type II superconductors [18,19] have been explained by D'Anna *et al.* within the critical-state model of Bean for a non- monotonous $j_c(T)$ [18], and our measurements of $\chi''(T)$ partially support this scenario also for $Nb_3Sn$.

Finally we briefly show that the peculiar double-peak structure in $\chi''(T)$ can also manifest itself in thermal data by the presence of an associated dissipated power according to Eq. (1). We have measured the corresponding heating power in a homebuilt calorimeter using an excitation field $\mu_0 H_{ac} \approx 6.5\,\text{mT}$ in $\mu_0 H = 6\,\text{T}$ for various frequencies $f$. The results of these measurements are shown in Fig. 4. As the magnetic susceptibility could not be measured in such a large excitation field, we have extrapolated the expected heating power from the measured $\chi_{meas}$-raw-data with $\mu_0 H_{ac} \approx 1.7\,\text{mT}$ and $f = 200$ Hz using Eq. (1), and we reach a reasonable agreement with the measured heating powers. It must be noted here that the strong variation of the magnetic-heating power with temperature in the peak-effect region ought to be considered in thermal experiments using a simultaneous "vortex-shaking" field in order to prevent artifacts in the resulting heat-capacity data.

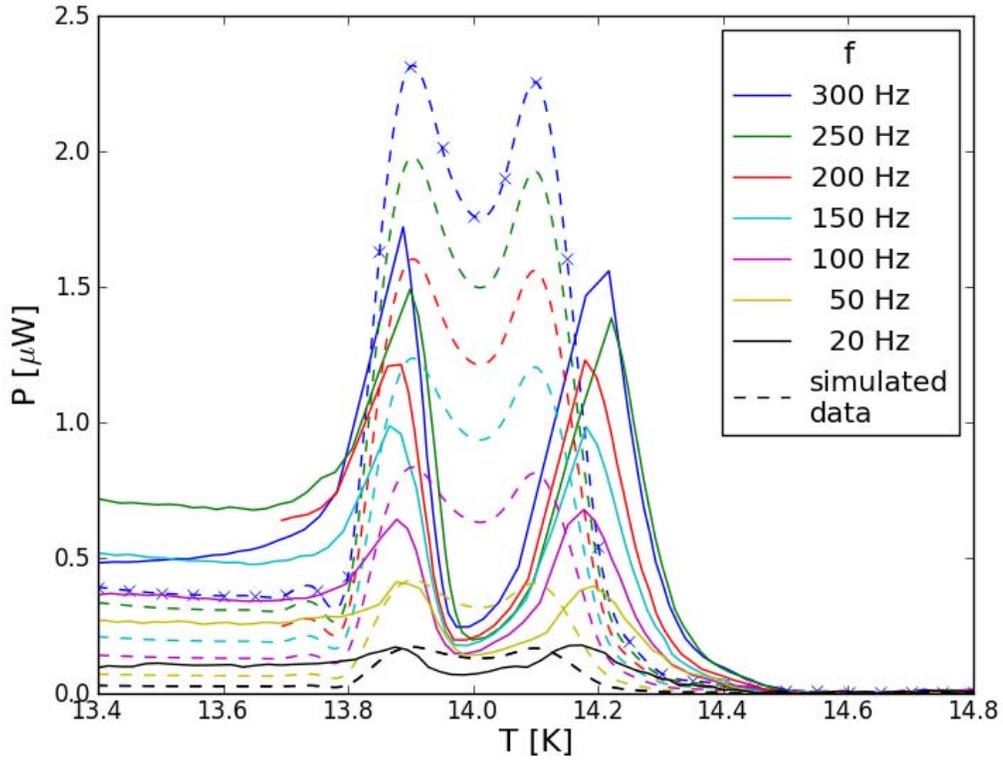

Fig. 4: Measured dissipative self-heating power in a $Nb_3Sn$ crystal exposed to a superimposed ac magnetic field with $\mu_0 H_{ac} \approx 6.5\,\text{mT}$ (solid lines) in $\mu_0 H = 6\,\text{T}$. Corresponding simulated data (dashed lines) were obtained by extrapolation from ac-susceptibility data taken with $\mu_0 H_{ac} \approx 1.7\,\text{mT}$ using a $B$-spline fit through the sampling points indicated by crosses in the simulated data for $f$ = 300 Hz, and using Eq. (1)

## IV. Conclusion

We have shown that the ac magnetic-susceptibility data of a $Nb_3Sn$ sample displaying a peak effect near $T_c$ can be well explained using a flux-creep model with a varying creep exponent $n$. While an $n$ ranging from 5 to 11 fits our data outside the peak-effect region reasonably well, a large exponent $n \approx 51$ which is close to Bean's critical-state model (i.e., $n \rightarrow \infty$) applies within the narrow peak-effect region near $T_c$. Since the value of the creep exponent does not alter the qualitative behavior of the $\chi''(\chi')$ relationship (with a single maximum at some intermediate value of $\chi'_m$ between −1 and 0), a single peak in the critical-current density always produces a single peak in the real part $\chi'(T)$ of the ac magnetic-susceptibility, which is accompanied by either a single peak in the imaginary part $\chi''(T)$ (for $|\chi'| \leq |\chi'_m|$) or a double peak (for $|\chi'| > |\chi'_m|$ in the peak effect region), respectively. Depending on the magnitude of the critical-current density in the peak-effect region, the sample geometry and the probing ac magnetic field, single or double peaks in $\chi''(T)$ may occur that are nevertheless manifestations of the same underlying physical phenomenon.

## Acknowledgements

This work was supported by the Schweizerische Nationalfonds zur Förderung der wissenschaftlichen Forschung, Grant. No. 20-131899.